# Optical band gap and associated band-tails in nanocrystalline AlN thin films grown by reactive IBSD at different substrate temperatures

Neha Sharma*, Shilpam Sharma, K. Prabakar, S. Amirthapandian, S. Ilango, S. Dash and

A. K. Tyagi

Material Science Group, Indira Gandhi Centre for Atomic Research

Kalpakkam 603102, TN, India

**Abstract**

AlN thin films have been grown on Si (100) substrates by reactive ion beam sputter deposition (IBSD) at different substrate temperatures varying from room temperature (RT) to 500$^o$C. Substrate temperature induced microstructural transition from amorphous at RT, nanocrystalline at 300$^o$C to microcrystalline at 400$^o$C has been observed by Transmission Electron Microscopy (TEM). Average surface roughness ($R_a$) and morphology has been explored by using Atomic Force Microscopy (AFM). UV-VIS spectrophotometry has been employed to probe the substrate temperature induced changes in optical band-gap ($E_g$) of grown thin films in reflectance mode. It was found that $E_g$ was increased from 5.08 to 5.21 eV as substrate temperature was increased from RT to 500$^o$C. Urbach energy tail ($E_u$) along with weak absorption tail (WAT) energy ($E_t$) have been estimated to account for the optical disorder which was found to decrease with associated increase in $E_g$.

## 1. Introduction

Aluminum nitride (AlN), a III-group nitride, has emerged as a potential candidate for the development of a number of devices in microelectronic, optoelectronic and micromechanical industries. Its unique optical, dielectrical and mechanical properties have enabled it for the fabrication of many technologically important devices [1-3]. As far as optical properties are concerned, a defect free single crystal AlN has a wide optical band-gap (6.2 eV), high refractive index (~2), low extinction coefficient (< 10$^{-3}$) and high thermal conductivity (285 W/m.K). This unique combination makes it competitive towards optoelectronic applications based on short-wavelength light emitting diodes, optical detectors and semiconductor lasers [4, 5]. It is highly insulating with a non-centrosymmetric lattice and exhibits nonvanishing second order susceptibility which makes it a promising material for applications in the field of nonlinear optics also [6]. However, the evolution of its optical properties in thin film form having short-range co-ordination or tiny crystallites with disordered boundaries in its microstructure is a matter of research. To this end, AlN thin films with different microstructures must be evaluated and characterized in detail to understand their optical response over a wide range of wavelength (λ). The knowledge of its microstructure dependent optical properties can then judiciously be used to quality them for numerous technological applications.

Optical band-gap of several group-III nitrides including their linear and nonlinear optical response have been extensively investigated both theoretically and experimentally during last decade [8-10]. In this context, there are several deposition techniques by which AlN thin films are grown on various substrates, such as metal organic chemical vapor deposition (MOCVD) [11], RF and DC reactive magnetron sputtering [12, 13], pulsed-DC reactive sputtering [14], molecular beam epitaxy (MBE) [15] and reactive ion beam sputter deposition (IBSD) [16, 17]. Among them, reactive IBSD is a technique by which smooth film surface and clear interface, a pre-requisite for device quality thin films, can be grown. Precise and independent control over incident ion beam energy and ion current further improves the technical importance of this particular technique [18]. At the same time, the most important aspect of RIBSD is the incorporation of defects in the near surface region by the energetic flux of reactive ions. This reactive ion induced disorder near the surface becomes the genesis of different film microstructure at different deposition parameters such as substrate temperature, assisted ion energy and current. Thus the fabrication of device quality AlN thin films with superior quality requires a better understanding of its optical band-gap as related to different microstructures at different deposition parameters. Although the available literature is replete with studies on reactive magnetron sputtered and laser ablated AlN thin films, the investigations on reactive

IBSD grown films and the role of substrate temperature on film microstructure and corresponding optical band-gap are rather sparse and motivates the current work.

In the present study, optical band-gap energy ($E_g$) of AlN thin films having different microstructures are investigated using a spectrophotometer. A comprehensive understanding on the microstructural evolution and its influence on the optical band-gap of AlN thin film is elucidated with an array of analytical techniques like transmission electron microscopy (TEM) and atomic force microscopy (AFM). The underlying dependence of microstructure on substrate temperature and the evolution of $E_g$ are addressed by assessing Urbach energy tail ($E_u$) and associated weak absorption tail energy ($E_t$).

## 2. Experimental

AlN thin films were deposited on 10 mm X 10 mm Si (100) substrates by reactive IBSD. The base pressure of the chamber was 4 X $10^{-7}$ mbar while working pressure was maintained at 4 X $10^{-4}$ mbar during deposition. Films were grown at different substrate temperatures varying from RT to 500°C in steps of 100°C. During deposition, the required metal atom flux was provided by sputtering an Al-target (99.999% pure) with an $Ar^+$ ion beam extracted from a 6 cm RF ion source with an energy of 500 eV. Concomitantly, an End-Hall type ion source was used to deliver an assistive reactive flux consisting of $N^+/N_2^+$ ions at 90 eV directly to the substrate surface.

The thickness of the films was measured ex-situ using a surface profilometer (Veeco, USA) and estimated to be ~114 nm for all the samples. Structural information of these films was obtained with high resolution TEM (Libra 200FE/HR-TEM). Selected area electron diffraction (SAED) patterns of these films were analyzed to determine the evolution of the crystallographic phases and average crystallite size (d) at different substrate temperatures. AFM (Ntegra Prima,

NT-MDT, Russia) was used to investigate the morphological features of the films over 1X1 $\mu m^2$ and 5X5 $\mu m^2$ area at multiple locations on a 10 mm X 10 mm substrate. Optical properties of these films were investigated using an UV-VIS spectrophotometer (Jasco V650, Japan). Same Si(100) substrate was used as reference for correcting the base line of all reflectance spectra collected in a wavelength ($\lambda$) range of 200 nm to 900 nm. Important optical parameters like $E_g$, $E_u$ and $E_t$ were calculated by analyzing the reflectance spectra obtained at normal incidence geometry.

## 3. Results and discussion

### 3.1 Film microstructure and Confirmation of AlN phase formation

Thin film microstructure and formation of AlN phase were investigated extensively by using HR-TEM. Microstructural information of the films is revealed in figure 1 which displays dark field images acquired by TEM at each substrate temperature. A careful analysis of these dark field images reveal that the films deposited at RT to 200°C substrate temperature possess a random atomic network and lack in long range coordination in its microstructure i.e. the films are amorphous in nature. This observation is further confirmed by the selected area electron diffraction (SAED) patterns shown in the inset of each image upto 200°C. However, for sample prepared at 300°C substrate temperature, onset of crystallization is noticed which advances with further increase in substrate temperature and evolves into a nanocrystalline microstructure at 400°C and 500°C. For clarity, few regions of interest in these dark field images are shown by bright rings. Nanocrystalline microstructure of these films was further analysed by acquiring SAED patterns at each substrate temperature from 300°C to 500°C shown in figure 2 along with an average crystallite size (d) distribution. The SAED pattern collected for the films deposited at 300°C substrate temperature shows a few weak Debye-Scherrer rings confirming the onset of crystallization with an average crystallite size (d) of 4.9 ± 0.3 nm in hexagonal phase of AlN. This process of crystallization continues at 400°C and reveals the presence of both hexagonal and cubic phases of AlN having d = 5.6 ± 0.3 nm with the appearance of many, relatively strong Debye-Scherrer rings. As the temperature of the substrate is further increased to 500°C, a number of strong Debye-Scherrer rings confirm the hexagonal phase of AlN to be the dominant constituent of film microstructure with d = 9.7 ± 0.2 nm. All these observations clearly establish simultaneous existence of different phases of AlN with varying fractions indicating the underlying structural transformations that strongly depend on substrate temperature during deposition. From RT to 200°C, films are predominantly amorphous in nature, become nanocrystalline (d<5 nm) at 300°C and evolve into microcrystalline (5 nm< d <50 nm) aggregates at 400°C and 500°C. These observations on film microstructure are listed in table.1

Surface morphology of these films was investigated using AFM which is shown in figure 3 at each substrate temperature. Surface morphology was found to be sensitive to the deposition temperature with the appearance of rapidly changing patterns and formation of cluster like structures, particularly at 500°C substrate temperature, which are highlighted with bright rings. Average surface roughness ($R_a$) obtained at several locations of 5x5 μm$^2$ areas were determined corresponding to each substrate temperature. A net increase in $R_a$ was noticed as the film microstructure transits from amorphous to nanocrystalline one with the increase in substrate temperature. These values of $R_a$ at each substrate temperature are listed in table.1. Also, composition of the films has also been studied by x-ray photoelectron spectroscopy (XPS) elsewhere [19] confirming the formation of AlN. Although, due to high affinity of aluminum towards oxygen, tiny domains of $Al_2O_3$, Al-O and Al-O-N complexes at the grain boundaries were also observed.

### 3.2 Calculation of Optical Band Gap Energy ($E_g$)

The fundamental absorption in a material takes place by excitonic process or band-to-band transition during interaction of photon with matter [20]. Herein, two types of transitions are possible at this fundamental edge; direct and indirect. Both these transitions involve interaction of photon with an electron in the valence band which subsequently gets excited into the conduction band across the fundamental band gap [21, 22]. However, in case of an indirect transition, simultaneous interaction with lattice vibration also takes place. The appearance of optical band-gap in thin films is greatly affected by its microstructural constituents like short-range co-ordination, nano and micro crystallites, defects etc.

The optical system under investigation was constituted of a thin and homogeneous film of AlN deposited on opaque Si (100) substrate. Thickness of the substrate was several orders of magnitude higher than that of the film which was measured to be 114 nm. Electromagnetic wave was made incident from the spectrophotometer to the film surface along a normal and reflectance spectra were collected. The reflectance spectra obtained for this study was analyzed using Kubelka-Munk (K-M) formalism wherein, the K-M function F(R) at any wave length is given by [23, 24]:

$$F(R) = \frac{(1-R)^2}{2R} \qquad (1)$$

Where, R is the absolute reflectance of the sample. F(R) values thus obtained were further converted into linear absorption coefficient ($\alpha$) by the following relation [25]:

$$\alpha = \frac{F(R)}{t} \qquad (2)$$

Where 't' is the thickness of the film which is ~114 nm for all the samples.

The *α* vs *hν* behaviour of AlN thin films is shown in figure 4. In figure 4(A), variation of α with hν for the samples deposited at RT is displayed which is found to be quite different from those deposited at higher substrate temperatures shown in figure 4(B). The important features of figure 4(A) and 4(B) are classified into three types of absorptions [26, 27]. First is the region-WAT which is attributed to the small optical absorption (α) governed by the optical transitions from one tail state to another tail state. These are called weak absorption tails (WAT). These localized tail states in amorphous semiconductors arise from defects generated disorder. Second is the region-U where α is controlled by the transitions from the localized tail states above the valence band to the extended states in the conduction band and/or from the extended states in the valence band to the localized states below the conduction band [26, 27]. In this region, the spectral dependence of α follows the Urbach rule. Third is the region-T which represents the range of α governed by the optical transitions from one extended state to another extended state. Most of the amorphous and nanocrystalline semiconductors follow Tauc's relation in this region. Thus $E_g$ is calculated by Tauc's plot in this region-T.

In the view of above discussion, the general form of the absorption coefficient *α* in region-T of figure 4(A) and 4(B) can be described by the Tauc's relation:

$$(\alpha h\nu)^{1/m} \sim A(h\nu - E_g) \qquad (3)$$

Where, $E_g$ is the optical band gap energy, A is a proportionality constant and *m* takes different values which correspond to different transitions [22]. Since AlN is known to be a direct band-gap material, value of *m* is taken to be ½ for this direct allowed transition. Thus, $E_g$ was determined by plotting '$(\alpha h\nu)^2$' as a function of photon energy '$h\nu$' and extrapolating the linear region of these plots to the x-axis as shown in figure 5. It is observed from the graphs that at RT, $E_g$ is found to be 5.08 eV. As the temperature of the substrate is increased to 100°C, $E_g$ increases subsequently to 5.13 eV. Maintaining this rising trend with further increase in substrate temperature $E_g$ becomes 5.19 eV, 5.20 eV and 5.21 eV respectively for 300°C, 400°C and 500°C. Although, the variation in $E_g$ is very small and can be approximated to the first decimal place i.e. 5.1 eV for samples grown from RT to 200°C and 5.2 eV for samples grown from 300°C to 500°C. But to understand the meV contribution from region-U and region-WAT, we are using the as calculated values. Keeping this in mind, we observe that $E_g$ experiences an increase in its values as the temperature of the substrate is increased from RT to 500°C during deposition and with the increase in crystalline fraction of AlN as well. This reduction in $E_g$ to ~5.2 eV of reactive IBSD grown AlN thin films from the reported value of 6.2 eV for single crystal AlN, can be elucidated by resorting to the concept of Urbach tails [29].

### 3.2 Determination of Urbach Energy ($E_u$) and weak absorption tail energy ($E_t$)

Generally, in optical absorption, near band edges, an electron from the top of the valence band gets excited into the bottom of the conduction band across the energy band gap [29]. During this transition process, if these electrons encounter disorder, it causes density of their states ρ(hν), where *hν* is the photon energy, tailing into the energy gap. This tail of ρ(hν) extending into the energy band gap is termed as Urbach tail. Consequently, absorption coefficient α(hν) also tails off in an exponential manner and the energy associated with this tail is referred to as Urbach energy and can be calculated by the following equation:

$$\alpha(h\nu) = \alpha_o \exp\left(\frac{h\nu}{E_u}\right) \qquad (4)$$

Where $\alpha_o$ is a constant, 'hν' is the photon energy and $E_u$ is the Urbach energy [30, 31]. In figure 4(A) and 4(B) region-U is the representative of this phenomena and corresponding range of α follows equation (4). The Urbach energy is estimated by plotting ln(α) vs. hν and fitting the linear portion of the curve with a straight line. The reciprocal of the slope of this linear region yields the value of $E_u$. The films deposited at RT are amorphous as these films have defects produced during reactive IBSD with negligible thermal activation for crystallization and Eu is estimated to be 310 meV. As the substrate temperature is raised to $100^oC$ and $200^oC$, $E_u$ reduces to 303 meV and 300 meV respectively. This indicates that the perpetual shuffling motion of atoms in randomly stacked polyhedra in an amorphous matrix becomes restricted with increase in substrate temperature. Such limited motion results in reduced number of vibrational energy levels for a given electronic state which diminishes the multiplicity in transition probability and contracts the Urbach tailing resulting in lower $E_u$. As the substrate temperature was increased to $300^oC$, a structural transition from amorphous to nanocrystalline state occurs. In this case, spatial variation with respect to bonding and local coordination, which is typical of short range order, takes place. This ensuing disorder density inherent to nanocrystalline state consequents a net value of $E_u$ = 293 meV. At $400^oC$, the structure of the film becomes microcrystalline which persists when substrate temperature further increased to $500^oC$. At these substrate temperatures, $E_u$ values are found to be nearly same, 290 meV and 289 meV, respectively. This behaviour of $E_u$ with increasing substrate temperature falls in-line with that of $E_g$ calculated in section 3.2. As the temperature of the substrate increases, film becomes more crystalline with increased band gap and reduced Urbach tail. These values of Eg and Eu are listed in table 2 at each substrate temperature and can be explained as following.

It is a well-known fact that ion beam assisted deposition generates near surface defects during deposition process. When an assistive ion with few tens of eV energy impinges on the surface of a thin film, a shallow penetration of the surface takes place and during the course of interaction with film-surface atoms, it transfers its energy to produce more number of knock-ons. Thus collision cascade propagates in the near surface region [32]. When the transferred energy is not sufficient to initiate a knock on, it gets absorbed in the form of lattice vibration and creates sub-surface defects such as vacancies, vacancy-interstitial pairs and antisites. Hence ion bombardment induced defects at the surface and sub-surface constitutes the source of disorder in the film [33]. It is generally well-recognized that AlN possess four residual point defects in its microstructure which are aluminum vacancy ($V_{Al}$), nitrogen vacancy ($V_N$), aluminum-antisites ($Al_N$) and nitrogen-antisites ($N_{Al}$) [34]. These defects introduce their own electronic states in the band gap and significantly alter the optical response of AlN. In figure 4(B), a much lower absorption peak at 2.8 eV to 3 eV is noticed which is not present in figure 4(A). This peak is likely owing to $V_N$ as proposed by Cox et al and Yim et al [35, 36]. These $V_N$ form a donor triplet just below the conduction band at 200 meV, 500 meV and 900 meV [35]. This feature indicates that for the films deposited at higher temperatures from $100^oC$ to $500^oC$, $V_N$ play a significant role in deciding $E_g$ but not for the films deposited at RT. Another point worth noticing here is that aluminum has very high affinity towards oxygen as compared to nitrogen. This becomes the gensis of oxygen related defects and defect complexes which are important intrinsic defects in AlN [37]. This oxygen in AlN matrix not only substitutes easily into nitrogen-sites but also forms $V_{Al} - nO_N$ (where n = 1, 2) type complexes and forms deep donor levels due to wide band gap of AlN. In *α vs. hν* curve shown in figure 4(A) and 4(B), a peak is noticed lying between 3.5 eV to 5 eV. This peak is assigned to oxygen absorption region as originally reported by Pastrnak and Roskovcova [38]. Thus we conclude that, all of the defects we mentioned above possibly exist in our AlN films but for the films grown at RT, only oxygen related defects and defect complexes dominate. While for the films grown at higher temperatures from $100^oC$ to $500^oC$, $V_N$, $O_N$ and $V_N-nO_N$ (where n = 1, 2) are the most probable and dominating defects.

In addition to above explained behaviour of α with hν, there exists a region-WAT in figure 4(A) and 4(B). Variation of α in this region is attributed to tail-to-tail state transitions as explained in section 3.2 and is governed by the following equation:

$$\alpha(h\nu) = \alpha_o \exp\left(\frac{h\nu}{E_t}\right) \qquad (5)$$

Where $E_t$ characterizes the equivalent band tail contribution arising from weak absorption tails [27]. Calculated values of $E_t$ are listed in table 2 and it is found that $E_t > E_u$ at each substrate temperature. As a matter of fact, with regard to the manifestation of optical properties of material, all kinds of disorders, irrespective of their origin, exhibit cumulatively additive property [39]. Thus to estimate the net contribution of structural and thermal disorder in film micostructure to account for the reduction in optical band gap of reactive IBSD grown AlN thin film, $E_u$ and $E_t$ can be added. If we plot $E_g$ vs. $E_u+E_t$, it is found to be linear as is shown in figure 6. When fitted with a straight line, its intercept gives the maximum possible value of $E_g = 6.24$ eV which can be considered to represent that value of $E_g$ when there is no disorder in thin film microstructure.

**Conclusion**

AlN thin films grown by reactive IBSD at different substrate temperatures were studied using TEM and AFM respectively for their structural and morphological features. Our results show that the films prepared at RT to 200°C were by and large amorphous in nature and became nanocrystalline for sample prepared at 300°C. When substrate temperature was maintained at 400°C during deposition, growth of microcrystallites was observed as a mixture of cubic and hexagonal phases of AlN. However, at 500°C it was found that the films were predominantly composed of hexagonal phase of AlN. These changes in microstructure as a function of substrate temperature had its corresponding impact on the optical band gap as revealed by the spectrophotometric studies. The origins of variations in optical band gap were found to stem from changes in microstructure at different substrate temperatures and give rise to significant band tailing into the band gap. The sub-surface defects such as $V_N$, $O_N$ and $V_{Al} - nO_N$, (where n = 1, 2) created by the assisted ion flux and preferable interaction of aluminum and oxygen during deposition of thin films, manifest as Urbach energy tail in the band-gap. The extent of increase in Urbach energy along with WAT energy can be used as a measure of disorder in the films. These observations indicate that a fine control over energy band-gap in combination with the temperature induced microstructure of AlN thin films can favorably be utilized to harness it in optical applications.

**Acknowledgement**

The authors would like to thank Mr. M. P. Janawadkar, Director, Materials Science Group for his encouragement and support. The authors acknowledge fruitful technical discussions with Dr. Awadhesh Mani, Dr. Arindam Das and Dr. V. Sivasubramanian.

**Figure Captions**

**Figure 1:** Dark field images acquired by TEM at each substrate temperature revealing structural transformations from room temperature (RT) to 500°C of AlN thin films. Selected area electron diffraction (SAED) patterns of the films prepared from RT to 200°C are shown in their in-set. While films deposited at 300°C to 500°C, crystalline region of interest are highlighted with bright rings.

**Figure 2:** Selected area electron diffraction (SAED) patterns along with average crystallite size (d) distribution for the samples prepared at 300°C to 500°C substrate temperature.

**Figure 3:** Surface morphology of AlN thin films deposited at different substrate temperatures acquired by atomic force microscopy (AFM). Some special morphological features were appeared at 500°C which are marked as dotted circles.

**Figure 4:** Absorption coefficient (α) as a function of photon energy (hν), (A) at RT, (B) 100°C to 500°C substrate temperatures

**Figure 5:** Tauc's plots to eastimate the optical band gap energy at each substrate tmperature

**Figure 6:** Comparative variation of $E_g$ with net $E_u + E_t$ . When fitted with a straight line, intercept of the linear fit (~ 6.24 eV) represents the optical band gap energy of the films when there is no disorder in their microstructure.

.

**Tables:**

Table 1. Microstructural details of the films extracted from TEM and AFM

| Substrate Temperature (°C) | d (in nm) | $R_a$ (in nm) |
|---|---|---|
| RT | Amorphous | 1.3 |
| 100 | Amorphous | 2.4 |
| 200 | Amorphous | 1.0 |
| 300 | 4.9±0.3 | 1.4 |
| 400 | 5.6±0.3 | 1.8 |
| 500 | 9.7±0.2 | 1.9 |

Table 2. Comparison of optical band gap energy ($E_g$), Urbach energy ($E_u$) and WAT energy ($E_t$) at different substrate temperatures

| Substrate Temperature (°C) | $E_g$ (in eV) | $E_u$ (in meV) | $E_t$ (in meV) |
|---|---|---|---|
| RT | 5.08 | 310 | 551 |
| 100 | 5.13 | 303 | 524 |
| 200 | 5.15 | 300 | 521 |
| 300 | 5.19 | 293 | 490 |
| 400 | 5.20 | 290 | 485 |
| 500 | 5.21 | 289 | 477 |

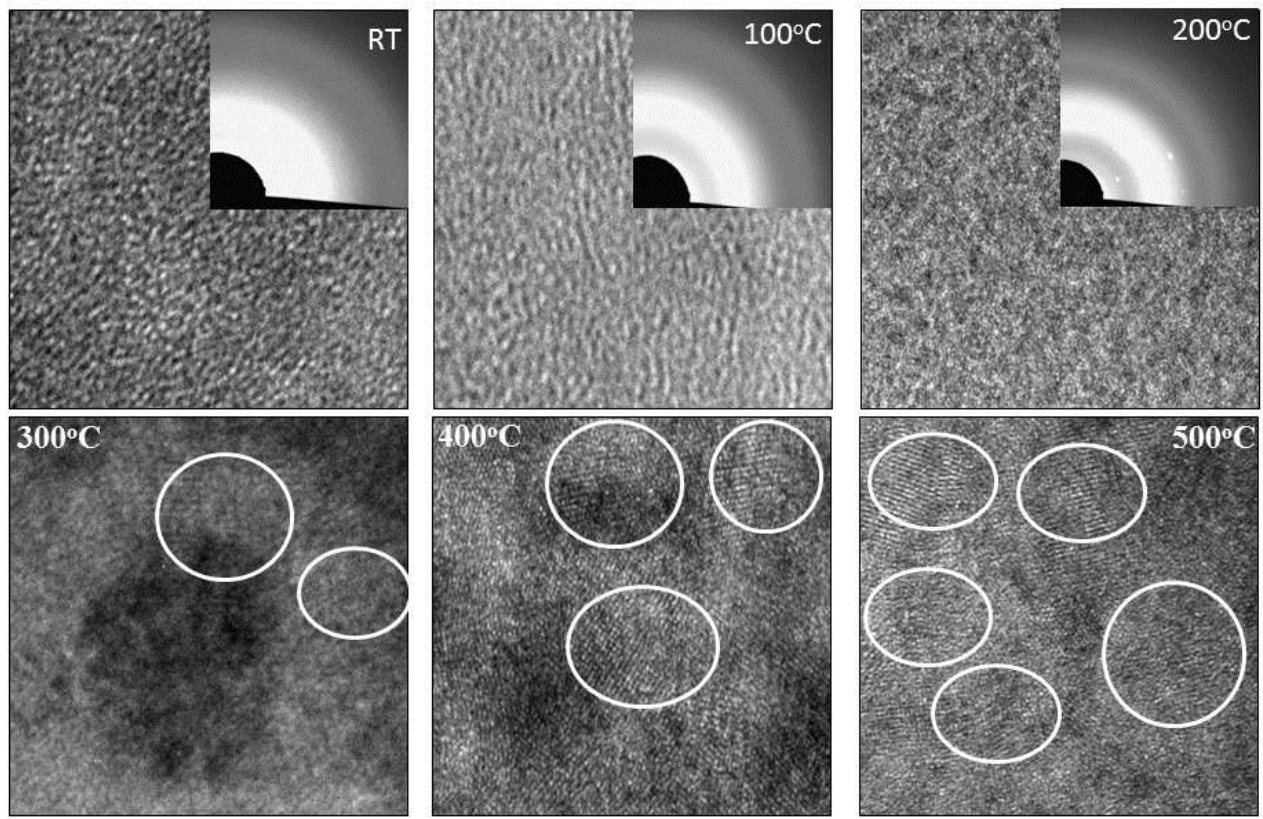
Figure - 1

**Figure - 2**

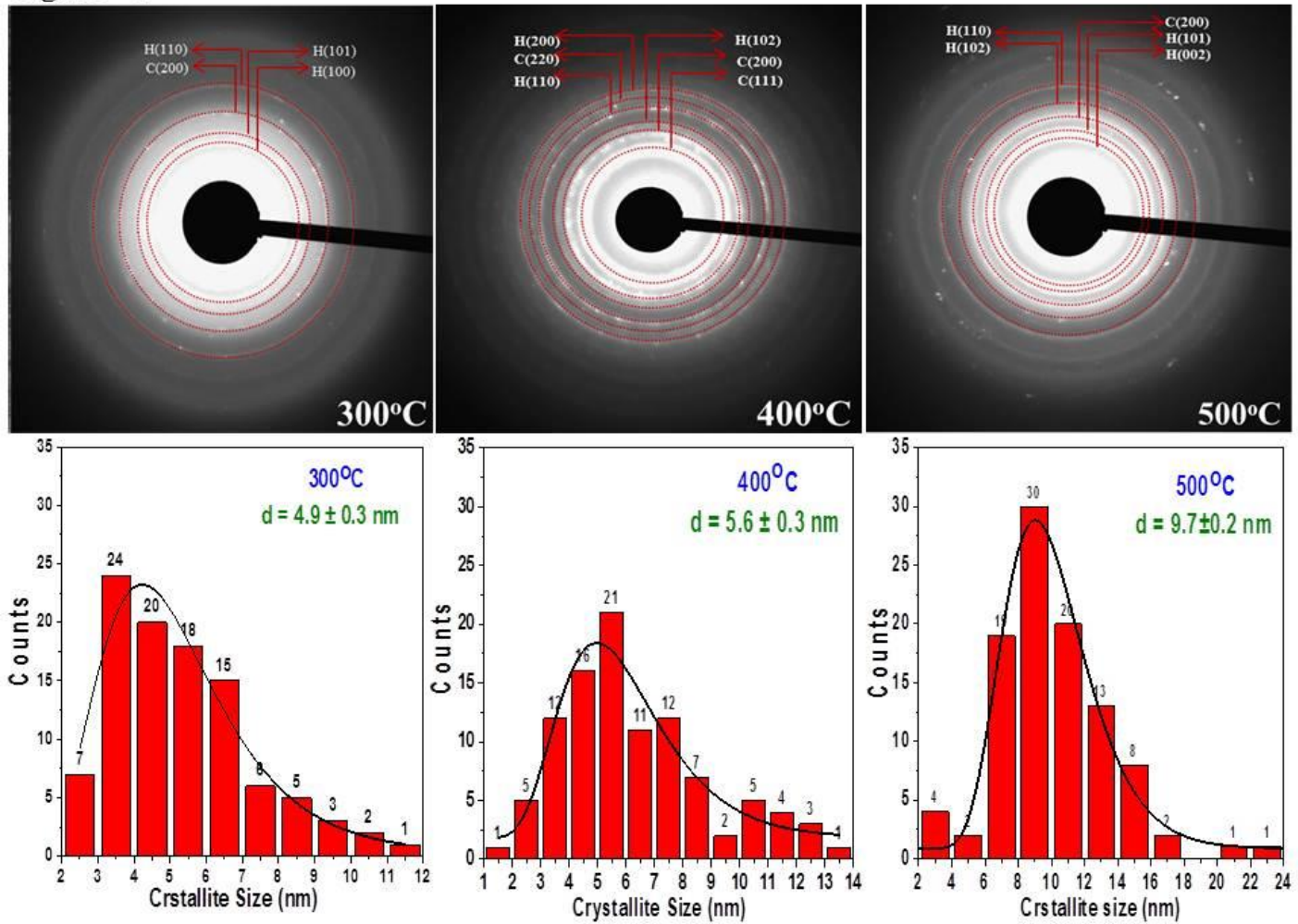

**Figure - 3**

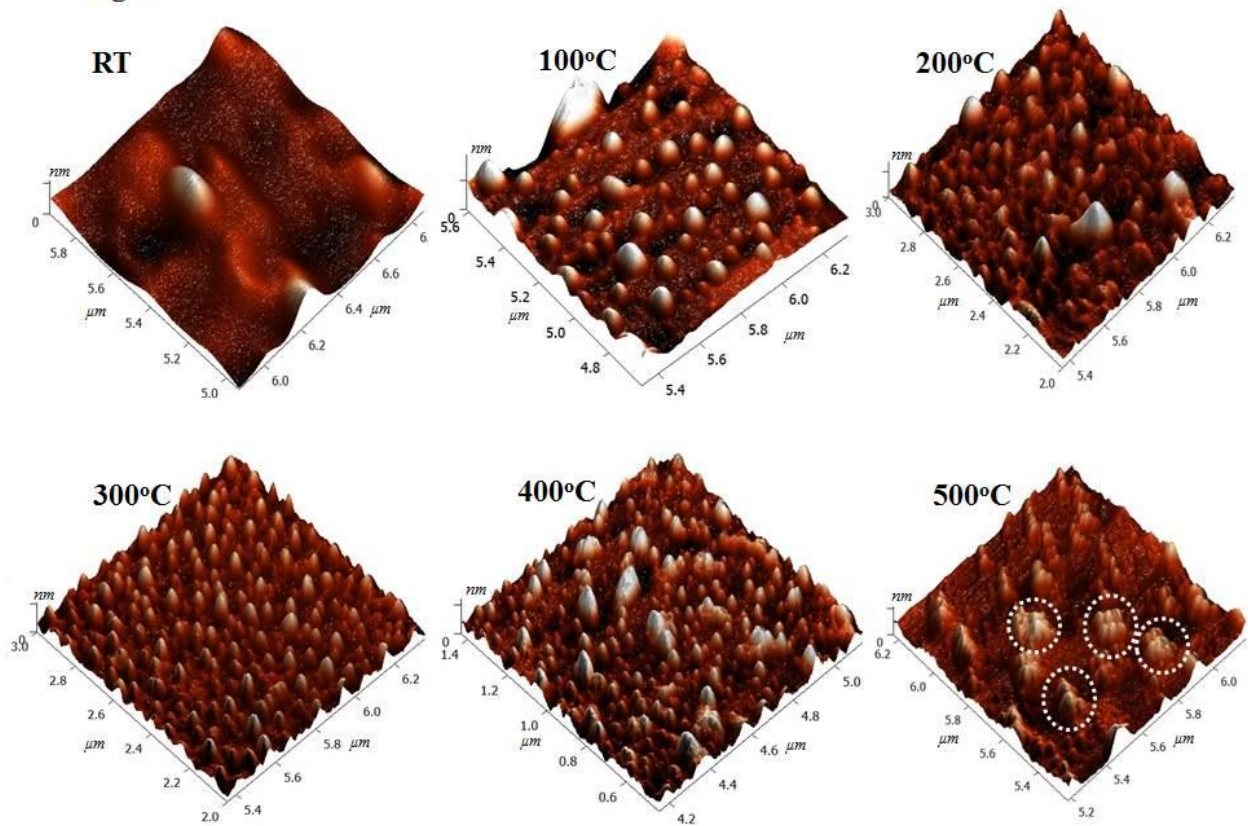

**Figure - 4**

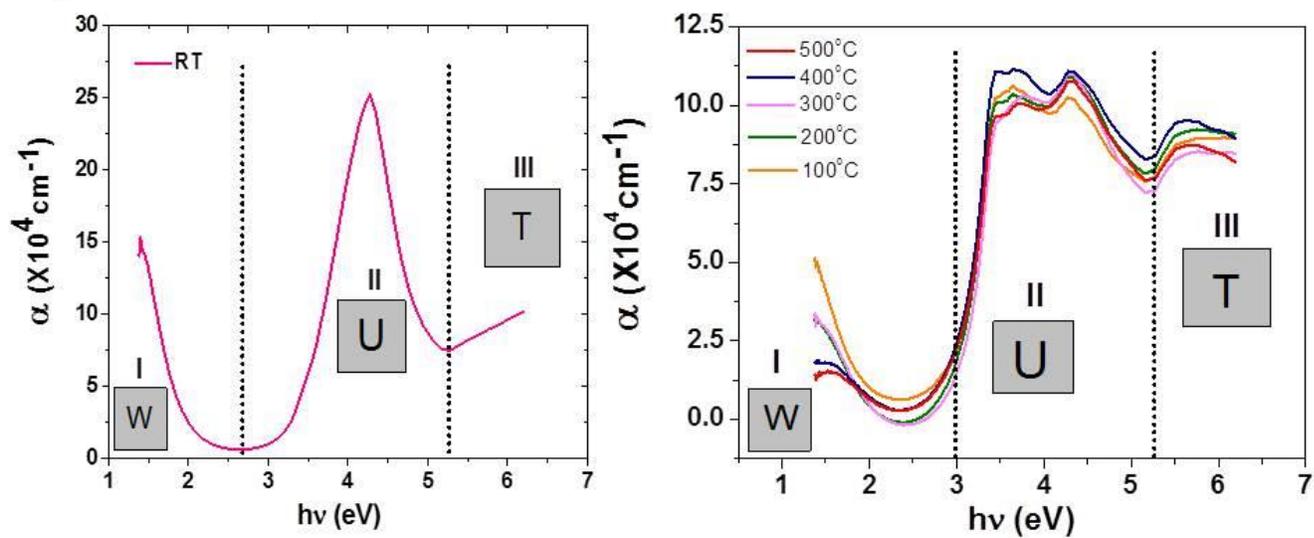

**Figure-5**

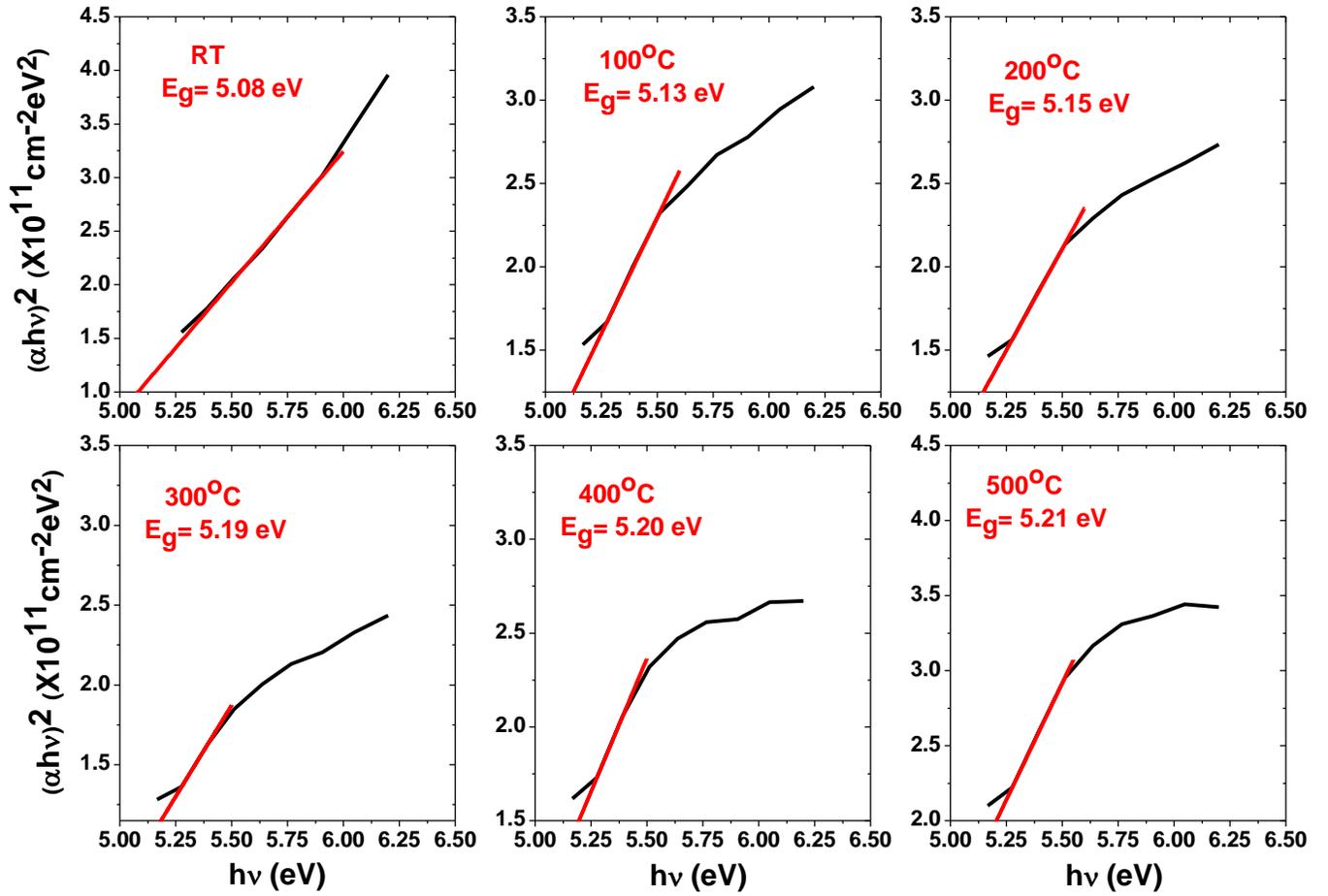

**Figure - 6**

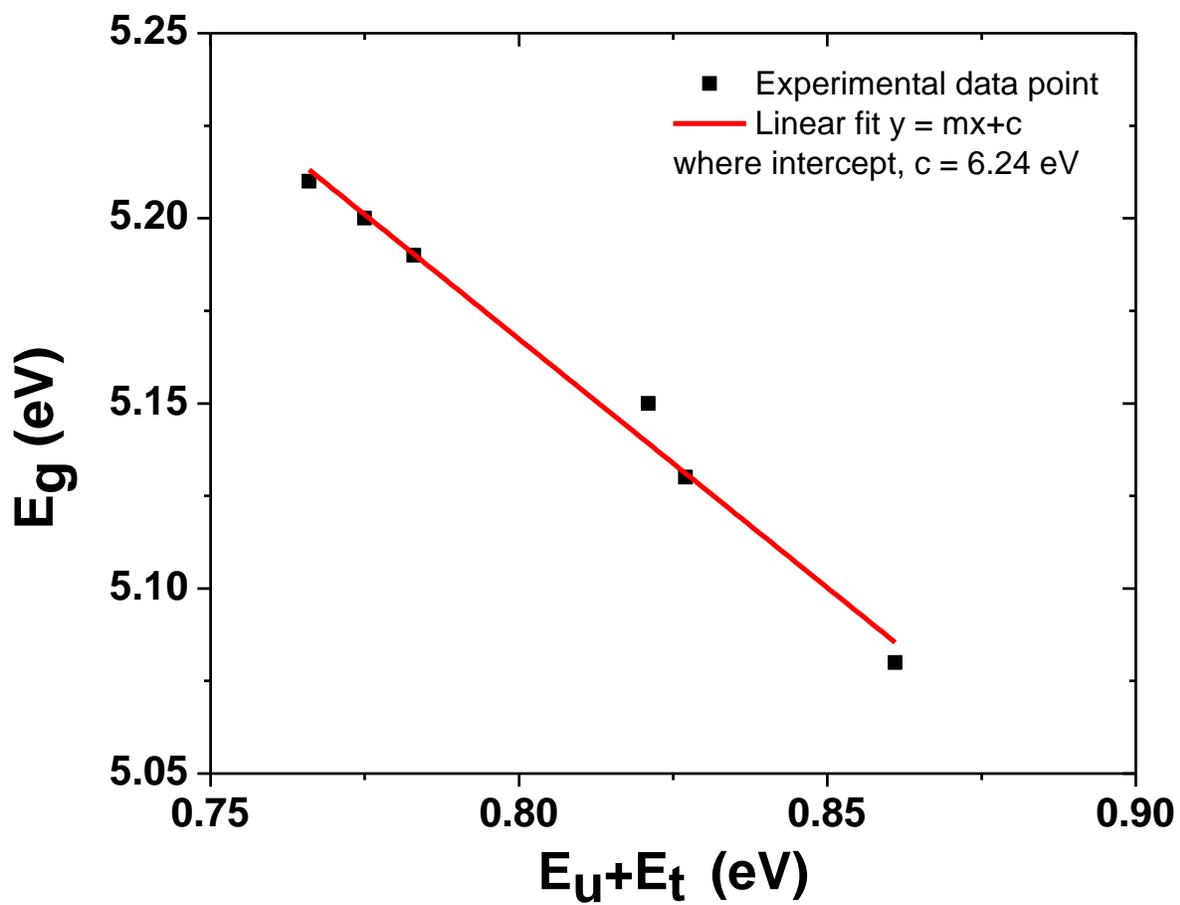